\input psfig.tex
\documentstyle[preprint,aps]{revtex}
\tighten
\begin{document}

\preprint{Imperial/TP/93-94/55}


\title{The Dynamics Of Vortex And Monopole Production By Quench
Induced Phase Separation} 

\author{A.J. Gill and R.J. Rivers}
\address{Blackett Laboratory, Imperial College, South Kensington,
London SW7 2BZ, U.K. \\ and \\ Isaac Newton Institute For Mathematical
Sciences, 20 Clarkson Road, Cambridge, CB3 0EH.}
\date{\today}
\maketitle



\begin{abstract}

Our understanding of the mechanism by which topological defects are
formed in symmetry breaking phase transitions has recently changed.
We examine the non-equilibrium dynamics of defect formation for
weakly-coupled 
global O(N) theories possessing vortices (strings) 
and monopoles.  It is seen that, as domains form and grow, defects are
swept along on 
their boundaries at a density of about one defect per coherence area
(strings) or per coherence volume (monopoles). 
\end{abstract}

\section{Introduction}

The formation of topological defects during symmetry breaking phase 
transitions is generic to many physical
systems. In particular we cite the vortices and monopoles of
superfluid $^{4}He$ and $^{3}He$ and the vortices (flux-tubes) of high- and low-$T_c$
superconductors. Similar defects, cosmic strings or monopoles,  most likely
appeared in the early universe at the GUT-scale phase transition.
All of these systems are described by some form of
quantum field theory and, due to the phase transition, their dynamics
is intrinsically non-equilibrium. They therefore provide a good means to
test non-equilibrium field-theory experimentally over a wide range of
energies. 

Roughly, the dynamics of defect formation proceeds as follows \cite{kibble1}.
From some initial state, which is not too far from thermal
equilibrium, some change in the bulk properties of the system, such as
pressure or volume, induces a phase transition. During this
transition, the scalar fields which describe the order parameter
fall from the false vacuum into the true vacuum,
choosing a point on the vacuum manifold at each
point in space, subject to the constraint that they must be
continuous and single-valued.
We shall limit ourselves to weakly first order or continuous
transitions, for which this collapse to the true vacuum 
occurs by spinodal decomposition or phase separation.
The resulting field configuration is one of
domains within each of which the scalar fields have relaxed to a constant
vacuum value.
If the theory permits defects, it will sometimes happen that the
requirements of 
continuity and single valuedness force the fields to remain in the false
vacuum between some of the domains. For example, in the case of
a complex scalar field producing vortices, the phase of the field may
change by an 
integer multiple of $2 \pi$ on going round a loop in space. This
requires at least one 
zero of the field within the loop, which signifies the presence of a
region of unbroken phase. Each zero has topological stability
and characterises a vortex passing through the
loop. When the phase transition is complete and there is no longer sufficient
thermal energy available for the field to fluctuate into the false
vacuum, the topological defects are frozen into the
field. From then on, the defect density alters almost entirely by
interactions of defects amongst themselves, rather than by
fluctuations in the fields, see for example \cite{tanmay}. 
 
The major question then, is what fixes the initial defect
density and the defect correlations? Only then can the subsequent
evolution of defect networks be determined with any accuracy.  
It was first argued that topological defects should be
frozen in at the Ginzberg temperature $T_{G}$ \cite{ginzburg}, the temperature above which
there is sufficient thermal energy available for the field to fluctuate
into the false vacuum without cost \cite{kibble1,ginzburg,ray}. If so, the defect number would
be strongly fluctuating above the Ginzberg temperature but
frozen in below it. In this case, the relevant scale for the initial
defect density would be the coherence length, $\xi(T_{G})$, of the
Higgs field (or fields) at the
Ginzberg temperature.  For example, in vortex production for the
$U(1)$ theory mentioned above, the initial vortex density
(i.e. the number of vortices passing through unit area) would be $\kappa /
\xi^2 (T_{G})$, where $\kappa$ is a constant of order unity.  Similarly, in
monopole production the monopole density would be expected to be
$\kappa / \xi^3 (T_{G})$, for similar $\kappa$.  Thereafter, defect
forces are assumed to take over.

Recently, however, more compelling pictures of the way in which the initial
density of topological defects is fixed have been proposed. 
While the mechanism outlined initially is almost certainly correct,
in general it is unlikely that  the Ginzberg temperature is relevant
to anything other than a thermally produced population of defects. 
For the cases of interest, for example an expanding universe, we expect
that, as the system is
driven from some initial thermal state towards the phase transition
there comes a 
point when the rate at which the transition is driven is too fast for 
the evolution of the field to keep up \cite{zurek1,zurek2}
The
transition may now be viewed as a quench and it is not clear that
either temperature or free energy mean anything at all. At this
point any
defects within the field are assumed to be frozen in until the transition is
complete. Upon completion, the field will try to return to thermal
equilibrium. At sufficiently low temperatures, however, the return to
equilibrium by thermal processes will be so slow that the evolution of
the initial defect density thus produced will be almost entirely by
interactions between the defects themselves.

Thus, in this scenario, the vortices cease
to be produced,
not at the Ginzberg temperature, but when the
scalar fields go out of equilibrium \footnote{Although thermal
equilibrium is mode dependent, this does not matter for the crude argument
repeated here.}. The relevant scale which determines the
defect density is the coherence length, $\xi(t)$ at this time,
and for some time onwards, rather than the coherence length at the 
Ginzberg temperature. 
The only remaining
uncertainty is how good an approxmiation it is to say that the defect
number is frozen into the field from the time when it first goes out
of equilibrium until
the time when it lies in the vacuum manifold almost everywhere and
its evolution can be viewed as the interaction of the defects. It is
this question which we address here. 

\section{The Model}

In the following we consider 
a class of theories where the broken and unbroken symmetries are {\it global}, thereby
guaranteeing that they pass through a second order transition. [Had
they passed through a strongly first order transition, the mechanism
for the transition, bubble nucleation, would lead to different
consequences from those outlined below, although it might still be a
good approximation to say that the defect density is frozen in when
the field first goes out of thermal equilibrium].
We assume that the change of symmetry is sufficiently rapid
that the fields are unable to respond immediately, but evolve by means
of phase separation or spinodal decomposition and domain formation.

We shall consider the simplest theory, one of $N$ massive
relativistic scalar fields $\phi_a$, where $a=1, \ldots , N$, in $D$
spatial dimensions, transforming as the fundamental representation of a
globally $O(N)$ invariant theory. Changes in the environment cause the
symmetry to be broken to $O(N-1)$ (i.e. as given by the generalised
'wine-bottle' potential) leading to a theory of one massive
Higgs boson and $N-1$ massless Goldstone bosons, with the vacuum manifold
$S^{N-1}$. Since the nth homotopy group $\Pi_n$ of the n-sphere is
$\Pi_n(S^n) = Z$, the group of integers, the theory possesses global
monopoles if $N=D$ and global strings if $N=D-1$. We are primarily
interested in $D=3$ dimensions, for which the $O(3)$ theory possesses
monopoles, and the $O(2)$ theory
possesses strings. [However, the vortex production in the $D=2$
Kosterlitz-Thouless transition has some interest, al though we shall
not pursue it here].

The transition is realised by the changing
environment inducing an explicit time-dependence in the field
parameters. Although we have the early universe in mind, 
we remain as simple as possible, in flat space-time
with the $\phi$-field action:-
\begin{displaymath}
S[\phi] = \int {\bf d^{D+1}x} \biggl (
\frac{1}{2} \partial_{\mu} \phi_a \partial^{\mu} \phi_a - \frac{1}{2}
m^2(t) \phi_a^2 - \frac{1}{4} \lambda(t) (\phi_a^2)^2
\biggr ).
\end{displaymath}
The t-dependence of $m^2(t)$ and $\lambda(t)$ is assumed given and
adjacent $O(N)$ indices are summed over. 

We wish to calculate
the evolution of the defect density during the fall from
the false vacuum to the true vacuum after a rapid quench from an
initial thermal state. The simplest 
assumption, which we shall adopt, is that the symmetry 
breaking occurs at time $t=t_0$, with the sign of $m^2(t)$ changing
from positive to negative at $t_0$. Further, after some short
period $\Delta t = t- t_0 >0$, $m^2 (t)$ and $\lambda (t)$ have relaxed to
their final values, denoted by $m^2$ and $\lambda$ respectively.
The field begins
to respond to the symmetry-breaking at $t=t_0$ but we assume that
its response time is greater than $\Delta t$, again ignoring any mode
dependence. 

To follow the evolution of the defect density during the fall off
the hill involves two problems.  The first is how to count the defects
and the second is how to follow the evolution of the quantum field.  We
take these in turn.

\section{Counting The Defect Density}

To calculate the defect density
requires knowledge of $p_t[\Phi]$, the
probability that, the measurement of 
the field $\phi(t,{\bf x}) = ( \phi_1, \phi_2, \ldots, \phi_N)$ would
yield the result $\Phi({\bf x}) = ( \Phi_1, \Phi_2, \ldots,
\Phi_N).$ This is obviously a consequence of both the initial conditions
and the subsequent dynamics. We follow Halperin \cite{halperin} in  adopting
a Gaussian distribution for
the field of the form:-
\begin{displaymath}
p_t[\Phi] = {\cal N} \exp \biggl (
- - \frac{1}{2} \int {\bf d^{D}x} \,{\bf d^{D}y} \, \, \Phi_a({\bf x})
K_{ab}({\bf x}-{\bf y};t) \Phi_b( {\bf y})
\biggr ),
\end{displaymath}
with $K_{ab} = \delta_{ab} K$ and ${\cal N}$ a
normalisation.  The circumstances under which a Gaussian is valid
will be examined later.  For weakly coupled theories 
we shall see  that, for short times after $t_0$ at least, a
Gaussian $p_t[\Phi]$ will occur.  If this is taken for granted it is
relatively straightforward to calculate the number density of defects.
Postponing the calculation of $K$  until then, we quote those of
Halperin's results that are relevant to us. 

Suppose that the field $\phi(t,{\bf x})$ takes the particular
value $\Phi({\bf x})$. We count the vortices by identifying them with
its zeroes.  The only way for a zero to occur with significant
probability is at the centre of a topological defect so, but for a
set of measure zero, all zeroes are topological defects.  
\footnote{This counting procedure alone gives no 
information about the length  distribution of the defects.}

Consider the $O(D)$ theory in $D$ spatial dimensions,
with global {\it monopoles}. Although less relevant than strings for
the early universe they are slightly easier to perform calculations
for. Almost everywhere, monopoles occur
at the zeroes of $\Phi({\bf x})$,
labelled ${\bf x_i}, i=1,2,\ldots$, at which the orientation
$\Phi({\bf x})/ |\Phi({\bf x}) |$ is ill-defined.
A topological winding
number $n_i=\pm 1$ can be associated with each zero ${\bf x_i}$ by the
rule:-
\begin{displaymath}
n_i = {\rm sgn} \det ( \partial_a \Phi_b) \biggr | _{ {\bf x}={\bf x_i}}. 
\end{displaymath}
Monopoles with higher winding number are understood as multiple zeroes
of $\Phi({\bf x})$ at which the $n_i$ are additive.
The {\it net} monopole density is then given by:-
\begin{displaymath}
\rho_{{\rm net}} ( {\bf x}) = \sum_i n_i  \, \delta ({\bf x} - {\bf x_i}).
\end{displaymath}
The volume integral of this gives the number of monopoles minus the
number of antimonopoles. The correlations of $\rho_{{\rm net}}$ give
us information on monopole-(anti)monopole correlations but, in the
first instance, we are interested in the cruder grand totals. 
The quantity of greater relevance to us,
is the {\it total} monopole density:- 
\begin{displaymath}
\rho({\bf x}) = \sum_i \delta({\bf x} - {\bf x_i}),
\end{displaymath}
whose volume integral gives the total number of monopoles {\it plus}
antimonopoles. in the volume of integration.

Now consider an ensemble of systems in which the fields $\Phi$ are
distributed according to $p_t[\Phi]$ at time $t$. Then, on average,
the total monopole density is:-
\begin{displaymath}
\langle \rho(t) \rangle = \biggl \langle \, \sum_i
\delta({\bf x} - {\bf x_i} ) \, \biggr 
\rangle _t,
\end{displaymath} 
where the triangular brackets denote averaging with respect to
$p_t[\Phi]$. That is:-
\begin{displaymath}
\langle F[\Phi] \rangle_t = \int {\cal D} \Phi \, F[\Phi] \, p_t[\Phi],
\end{displaymath}
with $p_t[\Phi]$ normalised so that $\int {\cal D} \Phi \, p_t[\Phi]
=1$. The translational invariance of the Gaussian kernel of the
probabililty density ensures that $\rho(t)$ is translationally invariant.

In terms of the fields $\Phi_a$, vanishing at ${\bf x_i}$, $\rho(t)$
can be re-expressed as:-
\begin{displaymath}
\langle \rho(t) \rangle = \langle \, \delta^D [\Phi_c({\bf x})] \, \,  |
\det \, (\partial_a \Phi_b({\bf x}) ) \, | \, \, \rangle_t.
\end{displaymath}
The second term in the brackets is just the Jacobian of the
transformation from ${\bf x}$ to $\Phi({\bf x})$. 

It follows from the Gaussian form of the probability density that 
the $\Phi_a$ individually and independently Gaussian
distributed with zero mean, as
\begin{displaymath}
\langle \, \Phi_a({\bf x}) \, \Phi_b ({\bf x}) \, \rangle _t =
\delta_{ab} W(|{\bf x} - {\bf y}| ;t),
\end{displaymath}
where $W(|{\bf x} - {\bf y}| ;t) = K^{-1}({\bf x} - {\bf y} ;t)$.
So also are the first derivatives of
the field $\partial_a \Phi_b$, which are independent of the field:-
\begin{displaymath}
\langle \Phi_c({\bf x}) \partial_a \Phi_b({\bf y}) \rangle_t = 0
\end{displaymath}
due to the fact that
$W$ is dependent only on the magnitude of ${\bf x} - {\bf y}$.

Thus, the total defect density may be separated into two independent
parts:-
\begin{displaymath}
\langle \rho(t) \rangle = \langle \, \delta^D [\Phi ({\bf x}) ] \,
\rangle_t \, \, \langle \, | \det ( \partial_a \Phi_b ) | \, \rangle _t.
\end{displaymath}
The first factor is easy to calculate, the second less so.

Consider first the delta-distribution factor:-
\begin{eqnarray}
\nonumber
\langle \delta[ \Phi ({\bf x_0}) ] \rangle_{t} &=& \int {\cal D} \Phi \,
\delta[ \Phi ({\bf x_0}) ] \, \exp \biggl \{ - \frac{1}{2}\int
d^{D}{\bf x} d^{D}{\bf y} 
\Phi({\bf x}) K({\bf x} - {\bf y};t) \Phi({\bf y}) \,
\biggr \} 
\\
\nonumber
&=& \int d \! \! \! / \alpha \int  {\cal D} \Phi \, 
\exp \biggl \{  i \alpha \Phi ( {\bf x_0} ) - \frac{1}{2}\int
d^{D}{\bf x} d^{D}{\bf y} 
\Phi({\bf x}) K({\bf x} - {\bf y};t) \Phi({\bf y}) \,
\biggr \},
\end{eqnarray}
where $O(N)$ indices and integrals over spatial variables have been
suppressed and $d \! \! \! / \alpha = d  \alpha / 2 \pi$. On defining
$\overline{\alpha} = \delta ({\bf x}-{\bf x_0} ) \alpha$, we find:-
\begin{eqnarray}
\nonumber
\langle \delta[ \Phi ({\bf x_0}) ] \rangle_{t} &=& \int  d \! \! \! /
\alpha \int {\cal D} \Phi 
\, \exp \biggl \{ \int d^{D}{\bf x}\, \biggl (i \overline{\alpha}({\bf x})
\Phi({\bf x})  - \frac{1}{2}\int
d^{D}{\bf x} d^{D}{\bf y}  \Phi({\bf x}) K({\bf x} - {\bf y};t) \Phi({\bf y})
\biggr ) \, 
\biggr \} \delta({\bf x}-{\bf x_0})
\\
\nonumber
&=& \int  d \! \! \! / \alpha \exp \biggl \{ - \frac{1}{2}\int
d^{D}{\bf x} d^{D}{\bf y} 
\overline{\alpha}({\bf x}) W(|{\bf x}-{\bf y}|;t)
\overline{\alpha}({\bf y}) \biggr \} \, 
\\
\nonumber
&=& \int  d \! \! \! / \alpha \exp \biggl \{ 
 - \frac{1}{2} \,\int
d^{D}{\bf x} d^{D}{\bf y}  \delta({\bf x}-{\bf x_0}) \alpha
W(|{\bf x}-{\bf x'}|;t) \alpha \delta({\bf x'}-{\bf x_0}) 
\, \,
\biggr \} 
\\
\nonumber
&=& \frac{1}{2 \pi} \biggl ( \frac{1}{\sqrt{K^{-1}}} \biggr ) ^2 =
\frac{1}{2 \pi \langle \Phi \Phi \rangle}.
=\frac{1}{2 \pi W(0;t)}
\end{eqnarray}

Consider now the second factor. Writing out the determinant
explicitly for $N=D=2$, and exploiting the fact that the field is
Gaussian, we have:-   
\begin{eqnarray}
\nonumber
\langle \,  | \, \det ( \partial_a \phi_b({\bf x})) \, | \,
\rangle_{t} ^2 &=& \biggl \langle \, \biggl [ \, \det ( \partial_a
\phi_b({\bf x}) ) \, \biggr ]^2 \, \biggr \rangle_{t}
\\
\nonumber
&=& \biggl \langle ( \partial_1 \phi_1 \partial_2 \phi_2
)^2 +(\partial_1 \phi_2 \partial_2 \phi_1)^2 - 2 \partial_1 \phi_2
\partial_2 \phi_1 \partial_1 \phi_1 \partial_2 \phi_2 \biggr \rangle_{t}
\end{eqnarray}
The first term may be factorised into a product of two Gaussian variables
and calculated as follows:- 
\begin{eqnarray}
\nonumber
\langle \, ( \partial_1 \phi_1 \partial_2 \phi_2 )^2 \, \rangle_{t} &=&
\langle \, (\partial_1 \phi_1 )^2 \, \rangle_{t} \, \langle \, (\partial_2
\phi_2 )^2 \,  \rangle_{t}
\\
\nonumber
&=& [ - \delta_{11} \partial_1 \partial_1 W({\bf x};t) ] \,  [ -
\delta_{22} \partial_2 \partial_2 W(|{\bf x}|;t) ] = [ \partial_1
\partial_1 W(|{\bf x}|;t) ]^2,
\end{eqnarray}
where $W(|{\bf x}|;t)$ is as before.
Fourier transforming the two-point function, we find:-
\begin{eqnarray} 
\nonumber
\langle \, ( \partial_1 \phi_1 \partial_2 \phi_2 )^2 \, \rangle_t &=&
\biggl ( \partial_1 \partial_1 \int e^{i {\bf k} . {\bf x} } 
\tilde{W}({\bf k};t) \,  {\bf d \! \! \! / ^3 k} \biggr )
\biggl ( \partial_2 \partial_2 \int e^{i {\bf k} . {\bf x} } 
\tilde{W}({\bf k};t) \,  {\bf d \! \! \! / ^3 k} \biggr ) 
\\
\nonumber
&=& \biggl ( 
\int k^2 \cos ^2(\theta) W({\bf k};t) \,  {\bf d \! \! \! / ^3 k}
\biggr )^2
\\
\nonumber
&=& [ \nabla^2 W(0;t) ]^2
\end{eqnarray}
A similar result applies for the second term whereas 
the third term vanishes. 
The final result is:-
\begin{displaymath}
\langle \rho(t) \rangle = C_N \Biggl | \frac{W''(0;t)} {W(0;t )}
\Biggr | ^{N/2}
\end{displaymath}
where the second derivative in the numerator is with respect to $x=|{\bf x}|$. $C_N$
is $1/ \pi^2$ for $N=D=3$ and $1/ 2 \pi$ for $N=D=2$ (had we
performed the calculation in $D=2$ dimensions), the difference
coming entirely from the determinant factor. 

Let us now consider the case of global strings in $D=3$ dimensions
that arise from the $O(2)$ theory. Strings are identified with
lines of zeroes of $\phi(t,{\bf x})= \Phi({\bf x})$ and the net vortex
density (vortices minus antivortices) in a plane perpendicular to the $i$-direction is:-
\begin{displaymath}
\rho_{{\rm net},i}({\bf x}) = \delta^2[\Phi({\bf x}) ] \, \epsilon_{ijk}
(\partial_j \Phi_1) ( \partial_k \Phi_2 ),
\end{displaymath} 
with obvious generalisations to $N=D-1$, for all $N$ in terms of the
Levi-Cevita symbol $\epsilon_{i_1,i_2, \ldots ,i_D}$.
As before, the total vortex density is of more immediate use. On a surface perpendicular to
the $i$-direction this is:-
\begin{displaymath}
\rho_{{\rm net},i}({\bf x}) = \delta^2[\Phi({\bf x}) ]
\, | \epsilon_{ijk} (\partial_j \Phi_1) ( \partial_k \Phi_2 ) |.
\end{displaymath}
in analogy with the monopole case.
The expectation value of this total density, when calculated as before
reproduces the same expression:-
\begin{displaymath}
\langle \rho(t) \rangle = C_N \Biggl | 
\frac{W''(0;t)} {W(0;t )}
\Biggr | ^{N/2},
\end{displaymath}
but for $N=D-1$. For the case of interest, $N=2$ and $C_2 =1/2 \pi$.
Thus,
whether we are concerned about global 
monopole or global string density, once we have calculated $K({\bf
x};t)$ we can find the total string density.
Similar results apply for the correlations between net densities,
which are important in determining the subsequent evolution of the
defect network.

\section{Evolution Of The Quantum Field}

During a symmetry breaking phase transition, the dynamics of the
quantum field is intrinsically non-equilibrium. The normal techniques
of equilibrium thermal field theory are therefore inapplicable. Out of
equilibrium, one typically proceeds using a functional Schr\"{o}dinger
equation or using the closed time path formalism of
Mahanthappa, Schwinger and Keldysh \cite{mahantapa,schwinger,keldysh}.
Here, we employ the latter, following closely the work of Boyanovsky,de Vega and coauthors \cite{devega,boyanovsky}.

Take $t=t_0$ as our starting time. Suppose that, at this time, the
system is in a pure state, in which the measurement of $\phi$ would
give $\Phi_0({\bf x})$. That is:-
\begin{displaymath}
\hat{\phi}(t_0,{\bf x}) | \Phi_0,t_0 \rangle = \Phi_0 | \Phi_0,t_0 \rangle.
\end{displaymath}
The probability $p_{t_f}[\Phi_f]$ that, at time $t_f>t_0$, the
measurement of $\phi$ will give the value $\Phi_f$ is $p_{t_f}[\Phi_f]
= |c_{f0}|^2$, where:-
\begin{displaymath}
c_{f0} = \int_{\phi(t_0) = \Phi_0}^{\phi(t_f) = \Phi_f} {\cal D} \phi
\, \exp \biggl \{ i S[\phi] \biggr \},
\end{displaymath}
in which ${\cal D} \phi = \prod_{a=1}^N {\cal D} \phi_a$ and spatial
labels have been suppressed. It follows that $p_{t_f}[\Phi_f]$ can be
written in the closed time-path form:-
\begin{displaymath}
p_{t_f}[\Phi_f] = \int_{\phi_{\pm}(t_0) = \Phi_0}^{\phi_{\pm}(t_f) =
\Phi_f} {\cal D} \phi_+  {\cal D} \phi_- 
\, \exp \biggl \{ i \biggl ( S[\phi_+]-S[\phi_-] \biggr ) \biggr \}.
\end{displaymath}
Instead of separately integrating $\phi_{\pm}$ along the time paths
$t_0 \leq t \leq t_f$, the integral can be interpreted as
time-ordering of a field $\phi$ along the closed path $C_+ \oplus C_-$
where $\phi =\phi_+$ on $C_+$ and $\phi= \phi_-$ on $C_-$.
\begin{figure}[t]
\centerline{\psfig{file=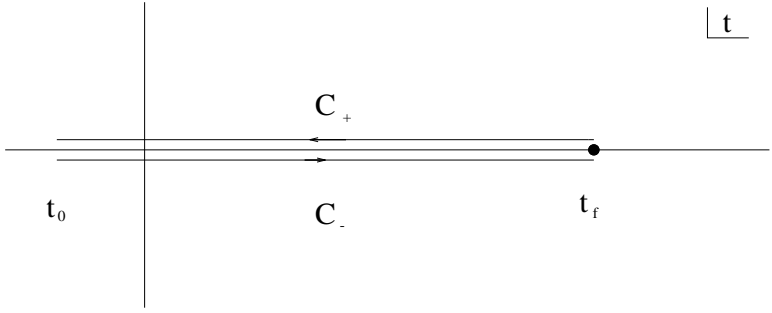,width=4in}}
\caption{The closed time path contour  $C_+ \oplus C_-$.}
\end{figure}
It is convenient to extend the contour from $t_f$ to $t= \infty$.
Either $\phi_+$ or $\phi_-$ is an equally good candidate for the
physical field, but we choose $\phi_+$:-
\begin{figure}[h]
\centerline{\psfig{file=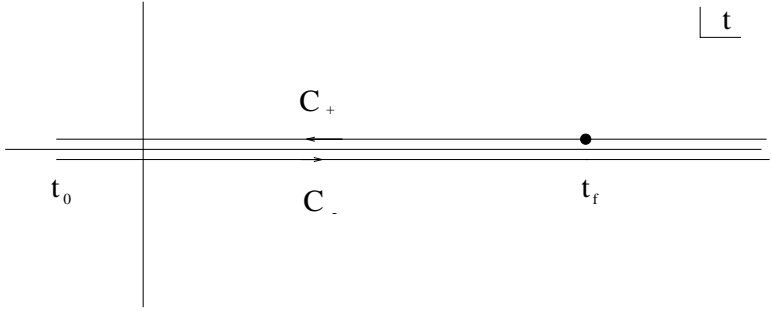,width=4in}}
\caption{Extending the integration contour.}
\end{figure}
With this choice and suitable normalisation, $p_{t_f}$ becomes:-
\begin{displaymath}
p_{t_f}[\Phi_f] = \int_{\phi_{\pm}(t_0) = \Phi_0} 
{\cal D} \phi_+  {\cal D} \phi_- \, \delta [ \phi_+(t) - \Phi_f ]
\, \exp \biggl \{ i \biggl ( S[\phi_+]-S[\phi_-] \biggr ) \biggr \},
\end{displaymath}
where $\delta [ \phi_+(t) - \Phi_f ]$ is a delta functional, imposing
the constraint $\phi_+(t,{\bf x}) = \Phi_f ({\bf x})$ for each ${\bf x}$.

The choice of a pure state at time $t_0$ is too simple to be of any
use. The one fixed condition is that we
begin in a symmetric state with $\langle \phi \rangle = 0$ at time
$t=t_0$. Otherwise, our ignorance is parametrised in the probability
distribution that at time $t_0$,
$\phi(t_0,{\bf x}) = \Phi({\bf x})$. 
If we allow for an initial probability
distribution $P_{t_0}[\Phi]$ then $p_{t_f}[\Phi_f]$ is generalised to:-
\begin{displaymath}
p_{t_f}[\Phi_f] = \int {\cal D} \Phi P_{t_0}[\Phi]
\int_{\phi_{\pm} (t_0) = \Phi} {\cal D} \phi_+  {\cal D}
\phi_- \, \delta [ \phi_+(t) - \Phi_f ] \, \exp \biggl \{ i \biggl (
S[\phi_+] - S[\phi_-] \biggr ) \biggr \}.
\end{displaymath}  
At this stage, we have to begin to make approximations. So that
$p_{t_f}[\Phi_f]$ shall be Gaussian, it is necessary to take
$P_{t_0}[\Phi]$ to be Gaussian also, with zero mean. All the cases
that we might wish to consider are encompassed in the assumption that
$\Phi$ is Boltzmann distributed at time $t_0$ at an effective
temperature of $T_0 = \beta_0^{-1}$ according to a quadratic Hamiltonian
$H_0[\Phi]$. That is:-
\begin{displaymath}
P_{t_0}[\Phi] = \langle \Phi,t_0 | e^{- \beta H_0} | \Phi,t_0 \rangle
= \int_{\phi_3(t_0) = \Phi = \phi_3(t_0-i \beta_0)} {\cal D} \phi_3
\exp \biggl \{ i S_0 [\phi_3] \biggr \},
\end{displaymath}
for a corresponding action $S_0[\phi_3]$, in which $\phi_3$ is taken
to be periodic in imaginary time with period $\beta_0$. We take
$S_0[\phi_3]$ to be quadratic in the $O(N)$ vector $\phi_3$ as:-
\begin{displaymath}
S_0[\phi_3] = \int {\bf d^{D+1}x} \biggl [ 
\frac{1}{2} (\partial_{\mu} \phi_{3 \, a} )(\partial^{\mu} \phi_{3 \,
a} )
- - \frac{1}{2} m_0^2 \phi_{3 \, a}^2
\biggr ].
\end{displaymath}
We stress that $m_0$ and $\beta_0$ parametrise our uncertainty in the
initial conditions. The choice $\beta_0 \rightarrow \infty$ corresponds
to choosing the $p_t[\Phi]$ to be determined by the ground state
functional of $H_0$, for example. Whatever, the effect is to give an
action $S_{3}[\phi]$ in which we are in thermal equilibrium for $t<t_0$ during
which period the mass $m(t)$ takes the constant value $m_0$ and, by
virtue of choosing a Gaussian initial distribution, $\lambda(t) = 0$
for $t<t_0$.

We now have the explicit form for $p_{t_f}[\Phi_f]$:-
\begin{eqnarray}
\nonumber
p_{t_f}[\Phi_f] &=& \int {\cal D} \Phi \int_{\phi_3(t_0) = \Phi =
\phi_3(t_0 - i \beta_0)} {\cal D} \phi_3 \,  e^{i S_0[\phi_3]}
\int_{\phi_{\pm}(t_0) = \Phi}
 {\cal D} \phi_+ \Phi {\cal D} \phi_- \, 
e^{i ( S[\phi_+] - S[\phi_-] ) } \delta [ \phi_+(t_f) - \Phi_f ]
\\
\nonumber
&=& \int_B {\cal D} \phi_3 {\cal D} \phi_+ {\cal D} \phi_- \, \exp \biggl \{ 
i S_0[\phi_3] + i ( S[\phi_+] - S[\phi_-] ) 
\biggr \} \, 
\delta [ \phi_+(t_f) - \Phi_f ],
\end{eqnarray}
where the boundary condition $B$ is $\phi_{\pm}(t_0) = \phi_3(t_0) =
\phi_3(t_0- i \beta_0)$. This can be written as the time ordering of a
single field:- 
\begin{displaymath}
p_{t_f} [ \Phi_f] = \int_B {\cal D} \phi \, e^{i S_C [\phi]} \, \delta [
\phi_+ (t_f) - \Phi_f ],
\end{displaymath}
along the contour $C=C_+ \oplus C_- \oplus C_3$, extended to include a
third imaginary leg, where $\phi$ takes the values $\phi_+$, $\phi_-$
and $\phi_3$ on $C_+$, $C_-$ and $C_3$ respectively, for which $S_C$
is $S[\phi_+]$, $S[\phi_-]$ and $S_0[\phi_3]$.
\begin{figure}
\centerline{\psfig{file=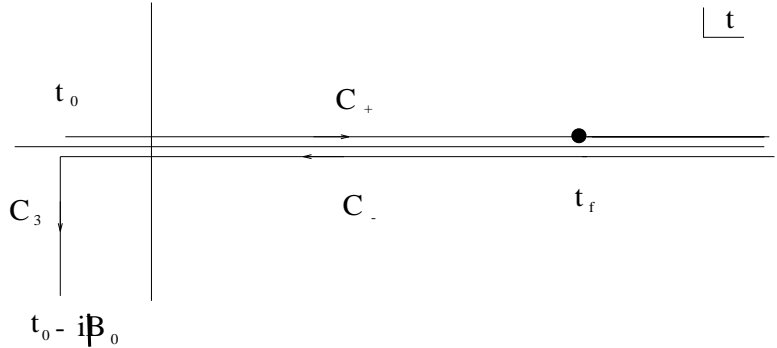,width=4in}}
\caption{A third imaginary leg}
\end{figure}
We stress again that
although $S_0[\phi]$ may look like the quadratic part of $S[\phi]$,
its role is solely to encode the initial distribution of configurations
$\Phi$ and need have nothing to do with the physical action.
Henceforth we drop the suffix $f$ on $\Phi_f$ and take the 
origin in time from which the evolution begins as $t_0 =0$.

We perform one final manoeuvre with $p_t[\Phi]$ before resorting to
further approximation. This will enable us to avoid an ill-defined
inversion of a two-point function later on. Consider the generating
functional:- 
\begin{displaymath}
Z[j_+,j_-,j_3] = \int_B {\cal D} \phi \, \exp \biggl \{ i S_C[\phi] +
i \int j \phi \biggr \},
\end{displaymath}
where $\int j \phi$ is a short notation for:-
\begin{displaymath}
\int j \phi \equiv \int_0^{\infty} dt \, \,  [ \, j_+(t) \phi_+(t) - j_-
\phi_-(t) \, ] \, + \int_0^{-i \beta} j_3(t) \phi_3(t) \, dt,
\end{displaymath}
omitting spatial arguments. Then introducing $\alpha_a({\bf x})$ where
$a=1, \ldots , N$, we find:-
\begin{eqnarray}
\nonumber
p_{t_f} [\Phi] &=& \int {\cal D} \alpha \int_B {\cal D } \phi \,\,
\exp \biggl \{i
S_C[\phi] \biggr \} \,\,
\exp \biggl \{ i\int {\bf d^4x} \alpha_a({\bf x}) [ \phi_+(t_f,{\bf
x}) - \Phi({\bf x}) ]_a \biggr \}
\\
\nonumber
&=& \int {\cal D} \alpha \, \, \exp \biggl \{ -i \int \alpha_a \Phi_a
\biggr \} \,Z[\overline{\alpha},0,0], 
\end{eqnarray}
where $\overline{\alpha}$ is the source $\overline{\alpha}(t,{\bf x}) =
\alpha({\bf x}) \delta (t-t_f)$. As with ${\cal D} \phi$, ${\cal D}
\alpha$ denotes $\prod_1^N {\cal D} \alpha_a$. 

We have seen that analytic progress can only be made insofar as
$p_t[\Phi]$ is itself Gaussian, requiring in turn that
$Z[\overline{\alpha},0,0]$ be Gaussian in the source
$\overline{\alpha}$. In order to treat the fall from the false
into the true vacuum, at best this means adopting a self-consistent
or variational approach i.e. a Hartree approximation or a large
N-expansion. \footnote{In the latter case, given the relationship
between $N$ and 
spatial dimension $D=N$, this corresponds to a large-dimension
expansion}.  However, 
if we limit ourselves to
small times $t$ then $p_t[\Phi]$ is genuinely Gaussian since the
field has not yet felt the upturn of the potential. That is, we may
treat the potential as an inverted parabola until the field begins to
probe beyond the spinodal point. The length of time
for which it is a good approximation to ignore the upturn of the
potential is greatest for weakly coupled theories which, for the
sake of calculation we assume, but physically, we
expect that if the defect counting approximation is going to fail, then
it will do so in the early part of the fall down the hill.
  
\section{Evolution Of The Defect Density}

The onset of the phase transition at time $t=0$ is characterised by
the instabilities of long wavelength fluctuations permitting the growth of
correlations. Although the initial
value of $\langle \phi \rangle$ over any volume is zero, 
the resulting phase separation or spinodal decomposition will lead to
domains of constant $\langle \phi \rangle$ whose boundaries will
evolve so that ultimately, the average value of $\phi$ in some finite
volume, will be non-zero. That is, the relativistic system has a
non-conserved order parameter. In this sense, the model considered
here is similar to those describing the $\lambda$ transition in liquid
helium or transitions in a superconductors.

Consider small amplitude fluctuations of $\phi_a$, at the top of the
parabolic potential hill described by $V(\phi) = \frac{1}{2} m^2 (t)
\phi_a^2$. At $t<0$, \, 
$m^2(t) > 0 $ and, for $t>0$,\, $m^2(t)<0$. However, by $t \approx \Delta
t$, \,$m^2(t)$ and $\lambda$ have achieved their final values, namely
$-\mu^2$ and $\lambda$. Long wavelength fluctuations, for which $|{\bf
k}|^2 < - m^2(t)$, begin to grow exponentially. If their growth rate
$\Gamma_k \approx \sqrt{-m^2(t) - |{\bf k}|^2}$ is much slower than
the rate of change of the environment which is causing the quench,
then those long wavelength modes are unable to track the quench. For
the case in point, this requires $m \Delta t \ll 1$. 
We take this to be the case. 
To
exemplify the growth of domains and the attendant dispersal of
defects, it is sufficient to take the idealised case, $\Delta t =0$ in
which the change of parameters at $t=0$ is instantaneous. That
is, $m^2(t)$ satisfies:-
\[ 
m^2(t) = \left \{
\begin{array}{ll}
m_0^2 >0 & \, \, \mbox{if $t<0$,} \\
 - \mu^2 <0 & \, \, \mbox{if $ t>0$}
\end{array} \right . 
\]
where for $t<0$, the field is in thermal equilibrium at inverse
temperature $\beta_0$. As for $\lambda(t)$, for $t<0$, it has already
been set to zero, so that $p_{t_0} [\Phi] $ be Gaussian. For small t,
when the amplitude of the field fluctuations is small, the field has
yet to experience the upturn of the potential and we can take
$\lambda(t) = 0$ then as well. At best, this can be valid until the
exponential growth $|\phi| \approx \mu e^{\mu t}$ in the amplitude
reaches the point of inflection $|\phi| \approx \mu / \sqrt{\lambda}$,
that is $\mu t \approx O(\ln (1/ \lambda) )$. The smaller the coupling
then, the longer this approximation is valid. 
As noted earlier, it should be possible to perform more sophisticated
calculations with the aim of evolving the defect density right through
the transition. For our present purposes, however, the small time or Gaussian
approximation is adequate.

We are now in a position to evaluate $p_t[\Phi]$, identify $K$ and
calculate the defect density accordingly. $S_C[\phi]$ becomes
$S_0[\phi_3]$ on segment $C_3$ so setting the boundary condition
$\phi_+(0,{\bf x}) = \phi_3(0,{\bf x} ) = \phi_3(-i \beta_0, {\bf x})$
and we have:-
\begin{displaymath}
S[\phi_+] = \int {\bf d^{D+1}x} \, \, \biggl [ \, \frac{1}{2}
(\partial_\mu \phi_a) 
(\partial^\mu \phi_a) + \frac{1}{2} \mu^2 \phi_a^2 \, \biggr ],
\end{displaymath}
on $C_+$. The Gaussian integrals can now be performed to give:-
\begin{displaymath}
p_t[\Phi] = \int {\cal D} \alpha \, \exp \biggl \{-i \int {\bf d^Dx}
\, \alpha_a  \Phi_a \biggr \}
\exp \biggl \{ \frac{i}{2} \int {\bf d^Dx} \,{\bf d^Dy} \, \alpha_a({\bf x})
G({\bf x} - {\bf y};t,t) \alpha_b({\bf y}) \biggr \}, 
\end{displaymath}
where  $G({\bf x} - {\bf y};t,t)$ is the equal time correlation, or
Wightman, function 
with thermal boundary conditions. Because of the time evolution
there is no time translation invariance in the double time label. 
As this is not simply
invertible, we leave the $\alpha$ integration unperformed. The form
is then a mnemonic reminding us that $K^{-1} = G$.

In fact, there is no need to integrate the $\alpha$s since from the
previous equation it follows that the characteristic functional
$\bigl \langle \exp \bigl \{ i \int J_a \Phi_a \bigr \} \bigr \rangle_t$ is directly
calculable as:- 
\begin{eqnarray}
\nonumber
\biggl \langle \exp \biggl \{ i \int j_a \Phi_a \biggr \} \biggr \rangle_t &=& \int {\cal D} \Phi
\, p_t[\Phi] \, \exp \biggl \{ i \int j_a \Phi_a \biggr \}
\\
\nonumber
&=& \exp \biggl \{ \frac{1}{2} \int {\bf d^4x} \, {\bf d^4y} \, j_a({\bf x}) G({\bf x} -
{\bf y} ; t ) j_a({\bf y}) \biggr \}.
\end{eqnarray}
Thus for example, the first factor in the monopole density $\rho(t)$
is:-
\begin{eqnarray}
\nonumber
\langle \, \delta ^D [ \Phi({\bf x}) ] \, \rangle_t &=& \Biggl \langle
\int dj \, \exp \bigl ( i \Phi_a({\bf x}) j_a \bigr ) \Biggr \rangle_{t} 
\\
\nonumber
&=& \int dj \, \exp \biggl \{ \frac {1}{2} j_a^2 G({\bf 0 };t,t) \biggr
\} = [ -iG({\bf 0};t,t)]^{-D/2},
\end{eqnarray}
with suitable normalisation, without having to invert $G({\bf
0};t)$. Thus, on identifying $-iG({\bf x};t,t)$ with $W({\bf x},t)$
as defined earlier, $\rho(t)$  becomes:-
\begin{displaymath}
\langle \rho(t) \rangle = C_N \Biggl | 
\frac { -iG''({\bf 0};t,t) } {-iG({\bf 0};t,t)}
\Biggr | ^{N/2}
\end{displaymath}
where $-iG({\bf x};t,t)$ has to be calculated from the equations of
motion, subject to the initial condtition.  

Details are given by Boyanovsky et al.,
\cite{boyanovsky} and we quote their results, which give $-iG({\bf x};t,t)$
as the real, positive quantity:-
\begin{eqnarray}
\nonumber
 -iG({\bf x};t,t) &=& \int \frac { d \! \! \! / ^D k }{2 \omega_<(k)}
\, e^{i {\bf k} . {\bf x} } \coth ( \beta_0 \omega_<(k) / 2)
\Biggl \{ 
\biggl [ 1+ A_k(\cosh(2W(k)t) - 1 ) \biggr ] \Theta(\mu^2 - |{\bf k }|^2)
\\
\nonumber 
&+& \biggl [ 1+ \alpha_k ( \cos(2\omega_>(k)t)-1 \biggr ] \Theta (
|{\bf k}|^2 - \mu^2)
\Biggr \} 
\end{eqnarray}
with:-
\begin{eqnarray}
\nonumber
\omega_<^2(k) = |{\bf k}|^2 + m_0^2
\\
\nonumber
\omega_>^2(k) = |{\bf k}|^2 - \mu^2
\\
\nonumber
W_<^2(k) = \mu^2 - |{\bf k}|^2
\\
\nonumber
A_k = \frac{1}{2} \Biggl (
1+ \frac{\omega_< ^2(k) }{W^2(k)} \Biggr ) 
\\
\nonumber
\alpha_k = \frac{1}{2} \Biggl ( 
1 - \frac{\omega_<^2(k) }{\omega_>^2(k)} \Biggr ).
\end{eqnarray}
The first term is the contribution of the unstable long wavelength
modes, which relax most quickly; 
the second is that of the short
wavelength stable modes which provide the noise. The first term will
dominate for large times and even though the approximation is only
valid for small times, there is a regime, for small couplings, in which
$t$ is large enough for $\cosh(2 \mu t) \approx \frac{1}{2} exp(2 \mu
t)$ and yet $\mu t$ is still smaller than the time $O(\ln 1/\lambda)$ at
which the fluctuations sample the deviation from a parabolic hill. In
these circumstances the integral at time t is dominated by a 
peak in the integrand $k^{D-1} e^{2W(k)t}$ at $k$ around $k_c$, where:-
\begin{displaymath}
t k_c^2 = \frac{(D-1)}{2} \mu \biggl ( 1 + O \biggl ( \frac {1}{\mu t}
\biggr ) \biggr ).
\end{displaymath}
The effect of changing $\beta_0$ is only visible in the $O(1/ \mu t)$
term. In the region $|{\bf x}| < \sqrt{ t/ \mu} $ the integral is
dominated by the saddle-point at $k_c$, to give:-

\begin{displaymath}
- - i G( {\bf x} ; t,t ) = W(x;t ) \approx W(0;t) \, \exp \Biggl
( \frac{- \mu x^2 }{ 8t} \Biggr ) \, {\rm sinc } \Biggl ( \frac{ |{\bf
x}| }{ \sqrt{t/\mu} } \Biggr ),
\end{displaymath}
for $D=3$, where:-
\begin{displaymath}
W(0;t) \approx C \frac {e^{2 \mu t} } {( \mu t) ^{3/2} },
\end{displaymath}  
for some C, which we don't need to know. The exponential growth
of $G({\bf 0 };t)$ in $t$ reflects the way the field amplitudes fall
off the hill $\langle \Phi \rangle = 0$. It is sufficient for our
purposes to retain $D=3$ only.

After symmetry breaking to $O(N-1)$ the mass of the Higgs is $m_H =
\sqrt{2} \mu$ with cold correlation length $\xi(0) = m_H^{-1}$. On
identifying $e^{- \mu x^2 / 8t}$ as $e^{-x^2 / \xi^2(t)}$ we interpret:-
\begin{displaymath}
\xi(t) = (8 \sqrt{2}) ^{1/2} \sqrt {\, t \, \xi(0) },
\end{displaymath}
as the size of Higgs field domains. This $t^{1/2}$ growth behaviour at
early times is characteristic of relativistic systems (with a double
time derivative) with a
non-conserved order parameter. 

To calculate the number density of defects at early times we have to
insert this expression for $-iG$ or $W$ into the equations derived
earlier.
Expanding $W(x;t)$ as:-
\begin{displaymath}
W(x;t) = W(0;t) \, \exp \biggl ( \frac{ - x^2}{\xi^2(t)}
\biggr ) \, \biggl ( \, 1 - \frac{4}{3} \frac{x^2}{\xi^2(t)} +
O \biggl ( \frac{x^2}{\xi^2(t)} \biggr ) \, \biggr ),
\end{displaymath}
and substituting in () we find:-
\begin{displaymath}
\langle \rho(t) \rangle = \frac{1}{\pi^2} \Biggl (
\frac{ \sqrt{14/3} } {\xi(t)} \Biggr )^3 \approx \frac{1.02}{\xi^3},
\end{displaymath}
for an $O(3)$ theory with monopoles in three dimensions and:-
\begin{displaymath}
\langle \rho(t) \rangle = \frac{1}{2 \pi} \Biggl (
\frac{ \sqrt{14/3} } {\xi(t)} \Biggr )^2 \approx \frac{0.74}{\xi^2},
\end{displaymath}
for an $O(2)$ theory with strings in three dimensions. 
The first observation is that the dependence of the density on time
$t$ is only through the correlation length $\xi (t)$.  As the
domains of coherent field form and expand, the interdefect distance
grows accordingly.  This we would interpret as the domains carrying the defects
along with them on their boundaries.  Secondly, there is roughly
one defect per coherence size,  a long held belief for whatever
mechanism.  However, in this case the density is exactly calculable. 

It is also possible to use Halperin's results to calculate
defect-defect correlation functions. For example, the monopole-monopole
correlation function on scales larger than a coherence length is found
to be:-
\begin{displaymath}
\langle \rho_{{\rm net}} ({\bf x}) \, \rho_{{\rm net}} ({\bf 0})
\rangle_t = \langle \rho(t) \rangle_t \, \delta({\bf x}) + g({\bf x})_t,
\end{displaymath}
where $g({\bf x})_t$ is a measure of teh screening of a monopole at
the origin. Explicit calculation yields:-
\begin{displaymath}
g({\bf x})_t = - \frac
{ 3 \sqrt{2} \exp ( - 3 x^2 / \xi^2 ) \sin^3 (2 \sqrt{2} x / \xi )}
{ 8 \pi^3 x \xi^5}.
\end{displaymath}

\section{Conclusions}

Under the conditions of a symmetry breaking phase transition, from
$O(N)$ to $O(N-1)$, which proceeds by a rapid quench, we have derived
expressions for the evolution of the defect density during the early
part of the fall from the false vacuum to the true vacuum. Our
results confirm that the defects are, indeed,
frozen into the field when it first goes out of thermal
equilibrium, at least for strongly coupled theories. 
Further, there is approximately one defect per correlation volume for
the time during which the approximation is valid.
The $t^{1/2}$ time-dependence of the correlation-length $\xi (t)$
that we have seen above is specific to a
non-conserved order parameter in a theory with a double time
derivative, but we expect the qualitative features to be similar for
all defect production by quenched symmetry breaking phase transitions.

Thus, generically, we expect the defect density to follow the
correlation length during the early part of the fall from the hill.
Further, we expect the correlation length to grow more during this
fall for weakly coupled theories, since the regime before which the
fields feel the upturn in potential and slow down is longer. For
very weakly coupled theories the
correlation length can grow significantly in the time interval
$t=O(ln(1/\lambda)$ available. In almost all physically
realistic scenarios except inflation, however, the coupling is not
small enough for this growth to be significant.  For more
strongly coupled theories we know less \footnote{although see the
self-consistent calulations of Boyanovsky et al.}. 
Whatever, it is quite reasonable to say that the defect
density is fixed at the time when the scalar field first goes out of
equilibrium.

\section*{Acknowledgements}

The authors would are grateful to T.S. Evans, M. Hindmarsh and T.W.B.
Kibble for useful conversations, particularly M. Hindmarsh for his
observations on Halperin and T.W.B. Kibble for his
understanding of the mechanism of defect formation. Both the authors
would like to thank the Isaac Newton Insitute for its hospitality and
Alasdair J. Gill would like to thank PPARC for its support.


\begin{thebibliography}{99}

\bibitem{kibble1} T.W.B. Kibble, J. Phys. {\bf A9}, 1387 (1976).

\bibitem{tanmay} T. Vachaspati and A. Vilenkin, Phys. Rev. {\bf D30},
2036 (1984).

\bibitem{ginzburg} V.I. Ginzburg, Fiz. Tverd. Tela {\bf 2}, 2031
(1960); Sov. Phys. Solid State {\bf 2}, 1826 (1961).

\bibitem{ray} M. Hindmarsh and R.J. Rivers, Nucl. Phys. {\bf 417},
506 (1994).

\bibitem{zurek1} W.H. Zurek, Nature {\bf 317}, 505 (1985). 
\bibitem{zurek2} W.H. Zurek, Acta Physica Polonica, {\bf B24}, 1301 (1993).

\bibitem{halperin} B.I. Halperin, Les Houches, Session XXXV 1980 NATO
ASI, editors Balian, Kl\'{e}man and Poirier.

\bibitem{mahantapa} K.T. Mahanthappa and P.M. Bakshi, J.M. Phys. {\bf
4}, 1; ibid, 12 (1963)

\bibitem{schwinger} J. Schwinger, J. Math. Phys. {\bf 2}, 407 (1961).

\bibitem{keldysh} L.V. Keldysh, Sov. Phys. JETP {\bf 20}, 1018 (1965).

\bibitem{devega} D. Boyanovsky, Da-Shin Lee and Anupam Singh, Phys. Rev. {\bf D48}, 800 (1993).

\bibitem{boyanovsky} D. Boyanovsky, H.J. de Vega, R. Holman. D-S.
Lee and A. Singh, preprint DOR-ER/40682-77

\end{thebibliography}
\end{document}